\documentclass[graybox]{svmult}

\usepackage{mathptmx}       
\usepackage{helvet}         
\usepackage{courier}        
\usepackage{type1cm}        
\usepackage{makeidx}         
\usepackage{graphicx}        
\usepackage{multicol}        
\usepackage[bottom]{footmisc}
\usepackage{epsfig,bbm,amsmath,amssymb,euscript,array,amsfonts}

\makeindex             

\begin{document}

\title*{Phenomenology of Unified Dark Matter models with fast transition}
\author{Alberto Rozas-Fern\'andez, Marco Bruni and Ruth Lazkoz}
\institute{Alberto Rozas-Fern\'andez \at ICG, University of Portsmouth,
UK, \email{alberto.rozas@port.ac.uk} \and Marco Bruni \at ICG, University of Portsmouth,
UK, \email{marco.bruni@port.ac.uk}
\and Ruth Lazkoz \at Dpto. de F\'\i sica Te\'orica, Universidad del Pa\'is
Vasco UPV/EHU, Apdo. 644, E-48080 Bilbao, Spain \email{ruth.lazkoz@ehu.es}}
%
%
\maketitle

\abstract*{}

\abstract{A fast transition  between a standard matter-like era and a late $\Lambda$CDM-like epoch generated by a single Unified Dark Matter component can explain the observed acceleration of the Universe. UDM models with a fast transition should be clearly distinguishable from $\Lambda$CDM (and
alternatives) through observations. Here we focus on a particularly simple model and analyse its viability by studying features of
the background model and properties of the adiabatic UDM perturbations.}

\section{Introduction}

 A possible framework explaining the acceleration of the Universe
is provided by models of Unified Dark Matter (UDM) where a single matter
component is supposed to source the acceleration and structure formation at the same time (see e.g. \cite{Bertacca:2010ct},
for a recent review).

UDM models with fast transition were introduced in \cite{Piattella:2009kt} and show interesting features
\cite{Piattella:2009kt,Bertacca:2010mt}.
The single UDM component must accelerate the Universe and provide acceptable perturbations which evolve in a scale-dependent fashion. In
view of testing models against observations this may become computationally  expensive. It is therefore essential to consider simple phenomenological models
of the fast-transition paradigm for which as much theoretical progress as possible can
be made from analytical calculations. This then can be used to increase the efficiency of numerical codes in dealing with these models.

 It turns out that the best receipt to proceed analytically is to prescribe
the evolution of the energy density of UDM.

\section{Generalities of UDM models}\label{sec:udmmodels}



\subsection{The background and the perturbations}

 We assume  a flat Friedmann-Lema\^{\i}tre-Robertson-Walker (FLRW) cosmology
where $w = p/\rho$
characterises the background of our UDM model.

We assume adiabatic perturbations. The squared Jeans wave number plays a crucial role in determining the viability of a UDM model, because of its
effect on perturbations, which is then revealed in observables such as the CMB and matter power spectrum
\cite{Pietrobon:2008js,Piattella:2009kt}. The explicit form of the Jeans wave number is \cite{Piattella:2009kt}
 \begin{multline}\label{kJ2analytic}
 k_{\rm J}^{2}  = \frac{3}{2}\rho a^{2} \frac{(1 + w)}{c_{\rm s}^2}\left|\frac{1}{2}(c_{\rm s}^2 - w) -
\rho\frac{dc_{\rm s}^2}{d\rho} + \frac{3(c_{\rm s}^2 - w)^2 - 2(c_{\rm s}^2 - w)}{6(1 + w)} + \frac{1}{3}\right|\;,
\end{multline} where $c_{\rm s}^2$ is the effective speed of sound.
So if we want an
analytic expression for $k_{\rm J}^{2}$ in order to obtain some
insight on the behaviour of perturbations in a given UDM model, we
need to be able to obtain analytic expressions for $\rho$, $p$,
$w$ and $c_{\rm s}^{2}$.


\section{Prescribing $\rho(a)$}\label{sec:3p}

Given a function (at least of class $C^{3}$) $\rho=\rho(a)$,  we can obtain the
following expressions for the quantities that enter into $k_{\rm J}^{2}$ (\ref{kJ2analytic}):
\begin{eqnarray}\label{varie1}
w & = & - \frac{a}{3}\, \frac{\rho'}{\rho}-1\;, \\
c_{\rm s}^{2} & = & - \frac{a}{3}\,\frac{\rho''}{\rho'} -\frac{4}{3}\;,
\label{varie2} \\
\frac{dc_{\rm s}^{2}}{d\rho}  & = & - \frac{1}{3\rho'^2}\, \left[
a \rho'''+\rho''-a \frac{\rho''^2}{\rho'}\right]\;. \label{varie3}
\end{eqnarray} where a prime indicates derivative with respect to $a$.


\section{Phenomenological UDM models with fast transition}\label{sec:tghmodel}

\subsection{A simple model for the background}

We introduce an ``affine" model
\cite{Pietrobon:2008js}:
\begin{equation}
\label{step}
\rho=  \rho_{\rm t}\left(\frac{a_{\rm t}}{a}\right)^{3} +\left[ \rho_{\Lambda} +(\rho_{\rm t}-\rho_{\Lambda})
\left(\frac{a_{\rm t}}{a}\right)^{3(1+\alpha)} -\rho_{\rm t}\left(\frac{a_{\rm t}}{a}\right)^{3} \right]
H_{t}(a-a_{\rm t})\;.
\end{equation}
$H_{\rm t}$ is compatible with having $c_{\rm s}^{2}>0$:
\begin{equation}
\label{Hc}
H_{\rm t}(a-a_{\rm t})=\frac{1}{2} + \frac{1}{\pi}\arctan(\beta(a-a_{\rm t})),
\end{equation} where the parameter $\beta$ represents the rapidity of the transition.
For $\alpha=0$, Eq.\ (\ref{step}) reduces to
\begin{equation}
\label{step1}
\rho=  \rho_{\rm t}\left(\frac{a_{\rm t}}{a}\right)^{3} +  \rho_{\Lambda} \left[ 1- \left(\frac{a_{\rm t}}{a}\right)^{3}
\right]
H_{t}(a-a_{\rm t})\;,
\end{equation}
 representing  a sudden transition to $\Lambda$CDM. In the following, we shall restrict our attention to this sub-class
of models.
Here $\rho_{\rm t}$ is the  energy scale at the transition,
$\rho_{\Lambda}$ is the effective cosmological constant and the redshift for
the
transition $z_{\rm t}=a_{\rm t}^{-1}-1$.

\section{The Jeans scale and the gravitational potential}\label{sec:perts}
\subsection{The Jeans wave number}

We require $k^{2} \ll k_{\rm J}^{2}$ for all scales of cosmological interest to which the linear perturbation
theory applies. A large $k_{\rm J}^{2}$ can be obtained not only when $c_{\rm s}^2
\to
0$, but when Eq.~(\ref{kJ2analytic}) is dominated by the
$\rho\; dc_{\rm s}^2/d\rho$
term.

Thus, viable adiabatic UDM models can be constructed which do not require $c_{\rm s}^2 \ll 1$  at all times if the
speed of sound goes through a rapid change, a fast transition period during which  $k_{\rm J}^{2}$ can remain large, in
the sense that $k^{2} \ll k_{\rm J}^{2}$.

In general $k_{\rm J}^{2}$
becomes larger, with a vanishingly small Jeans length (its inverse) before and after the transition. Although it becomes
vanishingly small for
extremely short times, the effects caused by its vanishing
are negligible (see the behaviour of the gravitational potential $\Phi$ below).

\subsection{The gravitational potential}

The equation that governs the behaviour of the gravitational potential $\Phi$ is \cite{Bertacca:2007cv}:

\begin{multline}\label{diff-eq_Phi}
\frac{d^2 \Phi(\textbf{k},a)}{d a^2}+\left(\frac{1}{\mathcal{H}} \frac{d \mathcal{H}}{d a} + \frac{4}{a}+ 3
\frac{c_{\rm s}^2}
{a}\right)\frac{d \Phi(\textbf{k},a)}{d a}+\\
\left[ \frac{2}{a \mathcal{H}}\frac{d \mathcal{H}}{d a} + \frac{1}{a^2}(1+3 c_{\rm s}^2)+\frac{c_{\rm s}^2 k^2}{a^2
\mathcal{H}^2}
\right] \Phi (\textbf{k},a) =0\;,
\end{multline}
where $\mathcal{H}= \frac{da}{d\eta}/a$ is the conformal time Hubble function. Also, $\mathcal{H}=aH$.

\begin{figure*}
\begin{center}
\includegraphics[width=0.325\columnwidth]{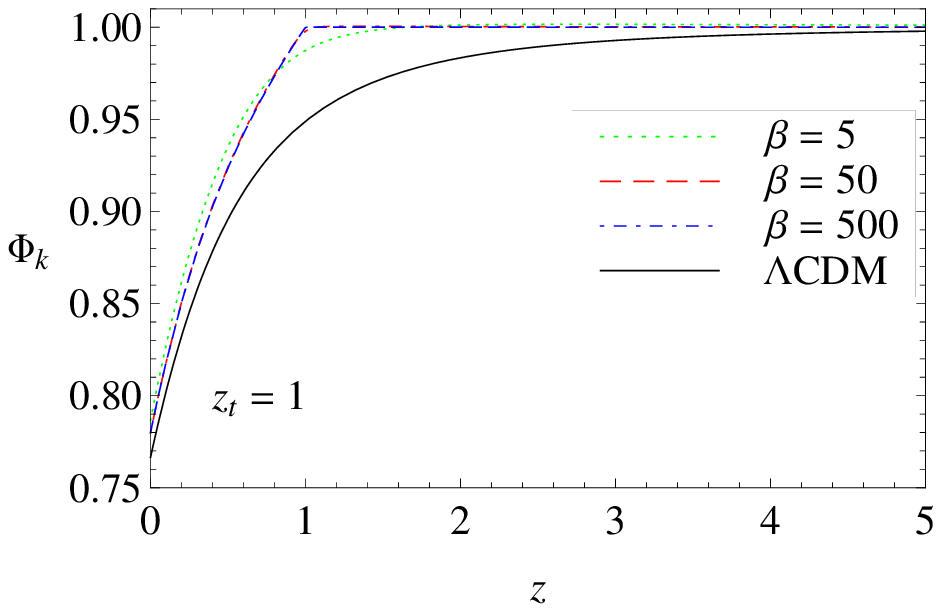}
\includegraphics[width=0.325\columnwidth]{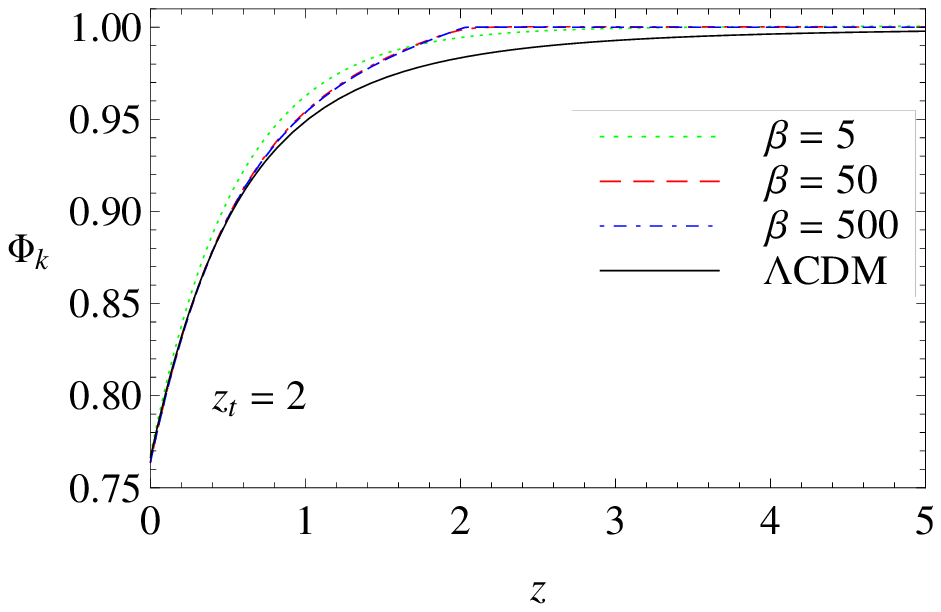}
\includegraphics[width=0.325\columnwidth]{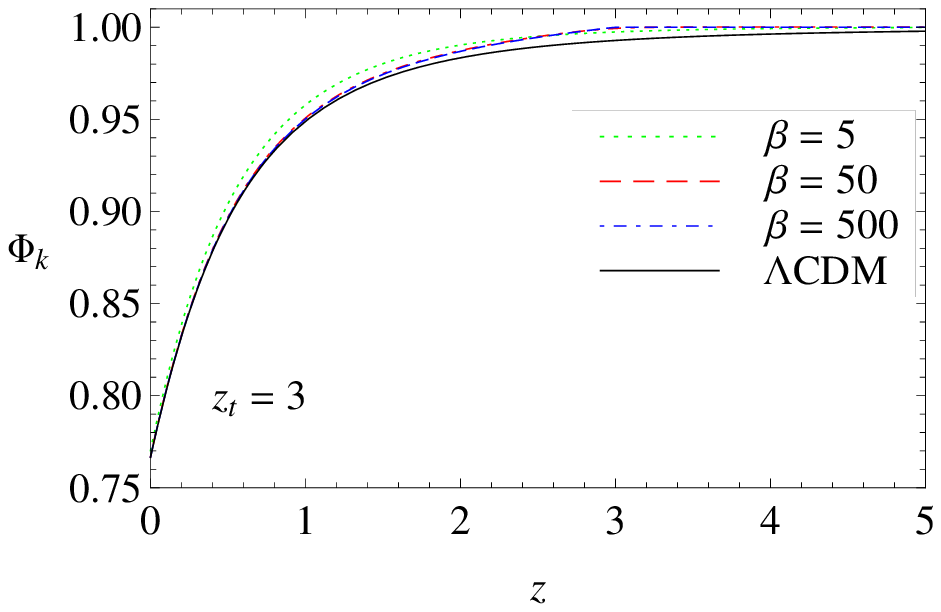}

\caption{Illustrative plots of the gravitational potential $\Phi(\textbf{k};z)$ as a function of the redshift $z$ for
$\Lambda$CDM and for our UDM model for $k=0.2$ $h$ Mpc$^{-1}$ and different values of $\beta$ and $z_{\rm t}$. The
black solid line corresponds to the gravitational potential in the $\Lambda$CDM model with $\Omega_{\Lambda,0} = 0.72$.}
\label{gpvsz}
\end{center}
\end{figure*}

From Fig.\ (1) we see that for an early enough fast transition with $\beta>500$ and $z_{\rm t} > 2$ our UDM model
should be compatible with observations. On the other hand, a study of the matter and CMB power spectra is needed to
study the viability of models with $10 \lesssim  \beta<500$, and those with $\beta>500$ and $z_{\rm t} < 2$.

\begin{acknowledgement}  MB is
supported by the STFC (grant no. ST/H002774/1), RL by the
Spanish Ministry of Economy and Competitiveness
through research projects FIS2010-15492 and Consolider
EPI CSD2010-00064, the University of the Basque Country UPV/EHU under
program UFI 11/55 and also by the ETORKOSMO special research action. ARF is supported by the `Fundaci\'on Ram\'on
Areces'.

\end{acknowledgement}


\begin{thebibliography}{99}
\bibitem{Bertacca:2010ct}
  D.~Bertacca, N.~Bartolo and S.~Matarrese,
  Adv.\ Astron.\  {\bf 2010} (2010) 904379
  [arXiv:1008.0614 [astro-ph.CO]].
\bibitem{Piattella:2009kt}
  O.~F.~Piattella, D.~Bertacca, M.~Bruni and D.~Pietrobon,
  JCAP {\bf 1001} (2010) 014
  [arXiv:0911.2664 [astro-ph.CO]].
\bibitem{Bertacca:2010mt}
  D.~Bertacca, M.~Bruni, O.~F.~Piattella and D.~Pietrobon,
  JCAP {\bf 1102} (2011) 018
  [arXiv:1011.6669 [astro-ph.CO]].
\bibitem{Pietrobon:2008js}
  D.~Pietrobon, A.~Balbi, M.~Bruni and C.~Quercellini,
  Phys.\ Rev.\ D {\bf 78} (2008) 083510
  [arXiv:0807.5077 [astro-ph]].
  \bibitem{Bertacca:2007cv}
  D.~Bertacca and N.~Bartolo,
  JCAP {\bf 0711} (2007) 026
  [arXiv:0707.4247 [astro-ph]].
\end{thebibliography}
\end{document}